\documentclass[5p]{elsarticle}

\usepackage{graphicx,amssymb,amsmath,color,hyperref}
\journal{Physica E}

\begin{document}

\begin{frontmatter}

\title{Waiting times of entangled electrons in normal-superconducting junctions}

\author{M. Albert}
\address{Universit\'e de Nice Sophia-Antipolis, INLN, CNRS, 06560 Valbonne, France}
\ead{mathias.albert@inln.cnrs.fr}

\author{D. Chevallier}
\address{Institut-Lorentz, Universiteit Leiden, P.O. Box 9506, 2300 RA Leiden, The Netherlands}

\author{P. Devillard}
\address{Aix Marseille Universit\'e, CNRS, CPT, UMR 7332, 13288 Marseille, France}
\address{Universit\'e de Toulon, CNRS, CPT, UMR 7332, 83957 La Garde, France}

\begin{abstract}
  We consider a normal-superconducting junction in order to investigate the effect of new physical ingredients on waiting times. First, we study the interplay between Andreev and specular scattering at the interface on the distribution of waiting times of electrons or holes separately. In that case the distribution is not altered dramatically compared to the case of a single quantum channel with a quantum point contact since the interface acts as an Andreev mirror for holes. We then consider a fully entangled state originating from spliting of Cooper pairs at the interface and demonstrate a significant enhancement of the probability to detect two consecutive electrons in a short time interval. Finally, we discuss the electronic waiting time distribution in the more realistic situation of partial entanglement. 
\end{abstract}

\begin{keyword}
Normal-superconducting junction \sep Andreev reflection \sep waiting time distribution \sep entanglement

\PACS 02.50.Ey \sep 72.70.+m \sep 73.23.Hk

\end{keyword}

\end{frontmatter}



\section{\label{sec:intro}Introduction}

Markus B\"uttiker was certainly one of the most influential scientists in the field of mesoscopic physics. Among all his important contributions, time in quantum mechanics has a peculiar flavor since it occupied his mind at the right beginning and at the end of his carrier. Intrigued at first by the traversal time of an electron through a tunnel barrier \cite{Buttiker1982,Buttiker1983}, he came back to this topic after the emergence of ``on-demand single electron sources'' \cite{Feve2007,Mahe2010,Dubois,Blumenthal2027,Pekola2008,Giazotto2011,Leicht2011,Hermelin2011,Fujiwara,Lansbergen,Tettamanzi}, which he greatly contributed to develop \cite{Buttiker1993,Buttiker1996,Buttiker2000,Buttiker2002,Buttiker2007,Buttiker2008,Albert2010,Parmentier2012,Haack2013,Buttiker2013}, via the concept of waiting time distribution (WTD) \cite{Albert2011,Albert12,Dasenbrook}. 

Charge transport at the nanoscale is known to be stochastic due to the quantum nature of particles \cite{Buttiker2000}. Therefore, going beyond the knowledge of average quantities, such as the average electronic current, appears to be unavoidable and extremely fruitful at the same time. A deep physical insight can indeed be inferred from the fluctuations of the signal and extracted from various observables. Noise \cite{Buttiker2000} and Full Counting Statistics (FCS) \cite{Levitov93,Levitov96,Nazarov2003}, namely the second moment of current fluctuations and the statistics of charges transferred during a long time interval, are among the most popular quantities and have been proved to be powerful tools. With the development of electron quantum optics \cite{Bocquillon2013} and the progress in single electron detection at high frequencies \cite{Mahe2010,Basset2012,Meunier2014,Reulet2014}, it is now relevant and possible to consider electron dynamics and time resolved quantities at quantum mechanical time scales (typically nano-seconds and below). Therefore, new theoretical tools have been developed to describe the current fluctuations at such time scales, such as finite frequency noise \cite{Mahe2010,Buttiker2000,Albert2010,Parmentier2012,Salo2006,Bednorz2010,Zamoun2012} and FCS \cite{Emary2007,Flindt2008,Hassler08,Marcos2010,Marcos2011,Ubbelohde2012}, Wigner functions \cite{Ferraro2013}, or the WTD \cite{Albert2011,Albert12,Dasenbrook,Welack2008,Brandes2008,FlindtThomas,AlbertDevillard,Rajabi2013,AlbertHaack14,Wang2014,Bjorn2014,Dasenbrook2015,Hofer2015}. The latter, describes the statistical distribution of time intervals between the detection of two electrons and therefore gives accurate information about correlations between subsequent electrons.

\begin{figure}
  \includegraphics[width=0.9\linewidth]{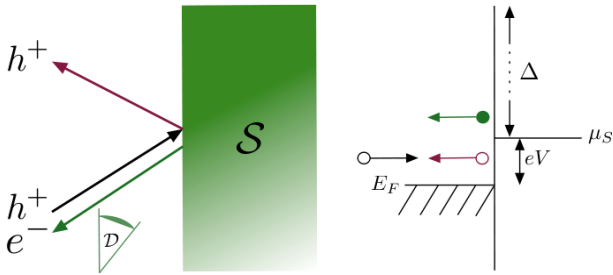}
  \caption{(Color online) Left: Schematic picture of a Normal-Superconducting junction. A hole approaching the interface from the normal part is either normally reflected or Andreev reflected back as an electron. A single electron detector is positioned to detect electrons from Andreev events. Right: energy diagram of the setup. The superconducting chemical potential $\mu_S$ is set to an energy $eV$ above the Fermi energy of the normal part and the gap $\Delta$ is much larger than the potential difference $eV$.} 
\label{fig:Fig1}
\end{figure}

The WTD has been studied for particularly simple systems like single and multiple electronic quantum channels connected to two normal leads via a Quantum Point Contact (QPC) \cite{Albert12,AlbertHaack14,Dasenbrook2015}, a quantum capacitor \cite{Albert2011,Hofer2015}, a double quantum dot \cite{Brandes2008,AlbertHaack14}, a train of Lorentzian pulses \cite{Dasenbrook,AlbertDevillard} or a quantum dot connected to a normal and a superconducting lead \cite{Rajabi2013,Braggio2011}, among others. In this paper we revisit the physics of Normal-Superconducting (NS) junction through the point of view of waiting times in order to illustrate the effect of superconducting correlations and entanglement \cite{LesovikMartinBlatter,Prada,Zheng,Braunecker} on their distribution. Indeed, as we will discuss later, such a system may emit entangled electrons in the normal part, and leads to interesting features in the WTD. 

The paper is organized as follows. In Sec. \ref{sec:model}, we describe the model used for the NS junction and the formalism needed for computing the WTD. In Sec. \ref{sec:onespin}, we discuss the effect of the transparency of the barrier (the energy dependence of the Andreev reflection) when the detection process is sensitive to only one electronic spin species and a certain range of energy. Section \ref{sec:twospins} is devoted to the effect of entanglement between spin up and spin down electrons emitted from the superconducting part, on the WTDs. We finally conclude and discuss some perspectives in Sec. \ref{sec:conclusion}. Moreover, for the sake of clarity, technical details are moved to the appendices. \ref{app:A} demonstrates the formal analogy between our setup and a single quantum channel conductor for a specific detection process whereas important steps for the numerical and analytical calculations of the WTD in the entangled case are explained in  \ref{app:B}. 

\section{\label{sec:model}Model}

One very important consequence of superconductivity is the existence of Andreev reflection. Such a phenomenon arises because the superconducting device cannot accommodate any single particle excitation with energy below the gap $\Delta$. Therefore, if a single particle like an electron or a hole flows from the normal part to the superconducting part with an energy below this threshold it can only be scattered back at the interface. However, there are now two possibilities. An electron (a hole) can be either normally reflected (specular reflection), that is to say, reflected as an electron (a hole) or converted to a hole (an electron). This is the so called Andreev reflection which originates from the fact that the incoming electron finds a partner to create a Cooper pair which can enter in the superconductor and leave a hole behind. 

To be more specific, the system of interest is a polarized NS junction (with an s-wave superconductor), at zero temperature, as presented on Fig. \ref{fig:Fig1}. The superconductor chemical potential $\mu_S$ is set to be at a potential $eV$ above the Fermi level $E_F$ of the normal metal. In such a situation, there is an incident hole, coming from the metal, that can be either normally reflected or Andreev reflected as an electron. Another way of picturing the Andreev effect is to think about the inverse configuration where a Cooper pair in the superconductor (at energy $\mu_S$ and zero momentum for an s-wave superconductor) splits at the interface and gives birth to an entangled pair of electrons. From now on, we will take $eV$ much smaller than the superconducting gap $\Delta$ in order to focus on this sub-gap phenomenon. This also has the benefit to make the Andreev time $t_A \equiv h/\Delta$ (the typical time needed for an Andreev event) much smaller than ${\overline \tau} \equiv h/(eV)$ (the typical time separation of two single particle wave packets emitted in the normal metal \cite{Albert12,Martin1992}). This allows us to assume that Andreev events are instantaneous and make use of scattering theory. In addition, this assumption allows one to linearize the dispersion relation around $\mu_S$ as $E(k)=\hbar v_F k$, with $E$ and $k$ measured from $\mu_S$ and its corresponding momentum (or around the Fermi level since $eV \ll E_F$ and $\mu_S$).

At the interface, the scattering is in general not perfect and both normal and Andreev reflection will play a role. In order to describe this effect, we use the standard Blonder-Tinkham-Klapwijk (BTK) model \cite{BTK} which has been widely used in the literature. The junction is modeled by a point-like barrier potential $U(x) = 2 Z E_F \lambda_F \delta(x)$, where $\lambda_F$ is the Fermi wavelength and $Z$ is a parameter measuring the strength of the barrier. It is then possible to compute the scattering matrix of this setup exactly and obtain the normal and Andreev transmission/reflection coefficients \cite{BTK,Duhot}. We do not reproduce these results in the present paper but give the corresponding numerical values of the coefficients when necessary.

Figure \ref{fig:Fig1} illustrates the scattering processes that we are now going to describe mathematically. The incident holes of energies $\mu_S - E$ lying between $E_F$ and $\mu_s$, arriving from the left and propagating to the right will be either normally reflected as holes of the same energies with amplitude $r_N$ or Andreev reflected as electrons of energies $\mu_S + E$ with amplitude $r_A$. The incoming scattering state is therefore a Slater determinant of holes of the form \cite{Samuelsson2003,Samuelsson,Zheng}
\begin{equation}
\vert \psi_{{\rm in}} \rangle\ =\prod^{eV}_{ E=0}c_{k(E),\uparrow}c_{k(E),\downarrow} \vert 0 \rangle\ ,
\end{equation}
where $\vert 0 \rangle\ $ stands for the filled Fermi sea up to $\mu_S$ in the normal part. However, in the electron language, this state is just the Fermi sea $\vert F \rangle\ $ filled up to $E_F$ instead of $\mu_S$. In the following, we will rather use the electronic picture to simplify the notation but both pictures are equivalent \cite{Zheng}. Due to scattering at the interface, the outgoing state is therefore a superposition of reflected holes, entangled electrons and non-entangled electrons \cite{Prada,Samuelsson2003,Samuelsson,Zheng, BTK,Duhot} 
\begin{equation}
  \begin{split}\label{eq:psiout}
    \vert \psi_{{\rm out}}\rangle\ &=\prod^{eV}_{E=0}(r^*_N(E)+r_A(E)c^\dagger_{k(E),\uparrow}c^\dagger_{-k(-E),\downarrow}) \\
    &\times (r_N(E)-r^*_A(E)c^\dagger_{k(E),\downarrow}c^\dagger_{-k(-E),\uparrow})\vert F\rangle\, .
  \end{split}
\end{equation}
Indeed, it is pretty straightforward to see that the previous equation, for a given energy, gives birth to three kinds of term with different levels of complexity. The terms $\vert F\rangle$ and $c^\dagger_{k(E),\uparrow}c^\dagger_{-k(-E),\downarrow}c^\dagger_{k(E),\downarrow}c^\dagger_{-k(-E),\uparrow} \vert F\rangle\ $ correspond to non-entangled contributions whereas $(c^\dagger_{k(E),\uparrow}c^\dagger_{-k(-E),\downarrow}-c^\dagger_{k(E),\downarrow}c^\dagger_{-k(-E),\uparrow})\vert F \rangle\ $ describes fully entangled electrons originating from the splitting of a Cooper pair at the interface. When Andreev reflection is absent ($r_a=0$), the Fermi Sea is unperturbed by the interface and nothing interesting happens. Counter-intuitively, perfect Andreev reflection does not lead to perfect entanglement. On the contrary, the state is a Slater determinant of non-entangled electrons and the NS junction acts as a conventionnal electron source \cite{Samuelsson}. It appears that the maximally entangled situation arises when Andreev and normal reflection probabilities are both equal to one half. Nevertheless, the WTD of a fully entangled state has never been studied to our knowledge and we will take the opportunity to study it in this paper before considering the general and more realistic state emitted at the interface.  

In order to conclude this section, we recall a few definitions about WTDs. As mentioned in the introduction, the waiting time $\tau$ is defined as the time delay between the detection of two single particles. Due to scattering and the quantum nature of particles, this time is a random variable, which distribution (the WTD) brings an elegant and instructive picture of the physics. For stationary systems, namely when there is no explicit time dependence, the WTD $\mathcal W(\tau)$ depends on $\tau$ only (and not on absolute time) and is closely related to the Idle Time Probability (ITP) $\Pi(\tau)$, the probability to detect no electron during a time interval $\tau$

\begin{equation}\label{eq:ITPWTD}
  \mathcal W(\tau)=\langle\tau \rangle\frac{d^2 \Pi(\tau)}{d\tau^2}\,,
\end{equation}
where $\langle\tau \rangle=-\left[\frac{d\Pi(\tau)}{d\tau}\right]^{-1}_{\tau=0}$ is the mean waiting time \cite{AlbertHaack14}.
To go further, we must now specify the detection procedure to compute the WTD. In what follows we will assume perfect single electron projective measurement but will consider two different situations as described below. Such perfect detection process is still theoretical and extremely challenging at very short time scales (nano-second and below) but recent experiments \cite{Mahe2010,Basset2012,Meunier2014,Reulet2014} are very promising about this issue. In order to compute the ITP, we regularize the scattering problem as usual \cite{AlbertHaack14}. We discretize energy, ranging from $0$ to $eV$, in $N$ slices and wave vectors as $k_n=\frac{2\pi n}{N}\frac{\vert eV\vert}{h v_F}$ and consider the limit $N\to \infty$ to mimic a stationary process.

\section{\label{sec:onespin}Andreev mirror for detecting WTD of holes}

The first application of this setup will be to measure the WTD of holes via the detection of Andreev reflected electrons or in other words the WTD of Andreev events \cite{Maisi}. A single electron detector is located at a position $x_0$ far away from the interface. In this section we assume that the detector measures only one spin orientation that we will choose upward for concreteness. Moreover, we make the additional assumption that it is only sensitive to energies above $\mu_S$, using a quantum dot for instance \cite{RecherLoss}. As a consequence, this allows us to avoid complications due to entanglement \cite{LesovikMartinBlatter,RecherLoss}, which will be the subject of the next section. Following \cite{Albert12,Dasenbrook,Levitov96}, the ITP is defined as 

\begin{equation}
  \Pi_\uparrow(\tau)  = \langle \psi_{{\rm out}} \vert :e^{- Q_{\uparrow,E>\mu_S}}:\vert \psi_{{\rm out}} \rangle
\end{equation}
with the explicit condition that only electrons with energies above $\mu_S$ contribute. Here $Q_{\uparrow}= \int_{x_0}^{x_0+v_F \tau} c_\uparrow^\dagger(x)\,c_\uparrow(x)\,dx$ and $:\cdots:$ stands for the normal ordering. In principle, the ITP cannot be reduced to a single determinant since the many body state is not a Slater determinant. However, according to the assumptions introduced above, only one term survives and the final result can be cast as a single determinant \cite{Albert12,Dasenbrook,Hassler08,AlbertHaack14}. Under these conditions, entanglement no longer plays a role and the problem thus boils down to a single quantum channel with energy dependent transmission \cite{AlbertHaack14}, where the role of the energy-dependent transparency $t(E)$ of the QPC is played here by $r_A(E)$, the Andreev reflection amplitude (see \ref{app:A} for details).
 
\begin{figure}
  \includegraphics[width=0.9\linewidth]{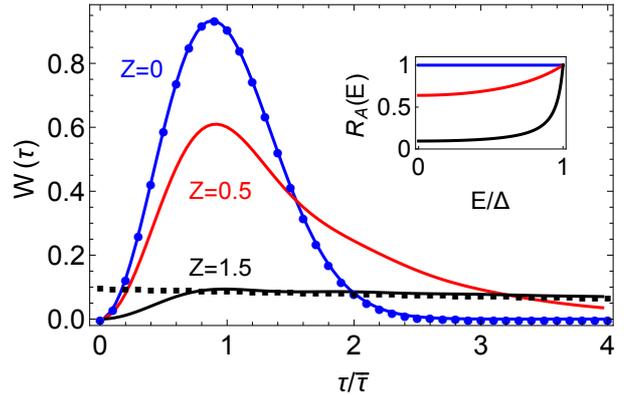}
  \caption{(Color online) WTD of Andreev events for various barrier transparencies. As $Z$ is increased, $R_A$ decreases from one to zero as shown in the inset. For $Z=0$, $R_A=1$ and the WTD is described by the Wigner distribution (blue dots, Eq. (\ref{eq:wigner})). Close to perfect specular reflection ($R_A\ll 1$) the distribution approaches an exponential law except for very small times (black squares). For intermediate $Z$, small oscillations with period $\overline \tau$ are superimposed to an exponential decay. Together, with the dip at $\tau=0$, they are manifestations of Pauli's exclusion principle \cite{Albert12}.}
  \label{fig:Fig2}
\end{figure}

In Fig. \ref{fig:Fig2}, we have plotted the spin up electronic WTD for different barriers strengths $Z=0$ (perfect Andreev reflection), $Z=0.5$ and $Z=1.5$ (strong barrier) as a function of $\tau/{\overline \tau}$. For information we show as an inset the corresponding Andreev reflection coefficient $R_A=|r_A|^2$ as a function of energy. However, the energy dependence is very weak since the energy window $eV$ is supposed to be much smaller than the superconducting gap $\Delta$. As expected, the WTD for $Z=0$ is approximately given by the Wigner surmise \cite{Albert12}

\begin{equation}\label{eq:wigner}
  \mathcal W_{\rm wd}(\tau)=\frac{32}{\pi^2}\, \frac{\tau^2}{\overline \tau^3} \exp\left[-\frac{4}{\pi} \left(\frac{\tau}{\overline \tau}\right)^2\right]\,.
\end{equation}
Indeed, in that case the train of free holes is perfectly converted at the interface into free electrons which are described by random matrix theory \cite{Albert12,Sutherland71}. As $Z$ is increased, the distribution is broadened since $R_A$ is no longer equal to one and therefore not all holes are converted into electrons. The situation is exactly equivalent to free electrons injected into a single quantum channel and partitioned by a quantum point contact with transmission probability $R_A$ due to the electron-hole symmetry in the system. The difference is only conceptual since here the detector is measuring indirectly the statistics of holes converted into electrons by the NS junction that acts as an Andreev mirror. Finally, for large $Z$, namely, small Andreev reflection, most of the holes are normally reflected and the detector collects rare events which are almost uncorrelated and the WTD is exponential (except for very short times). Indeed, following \cite{AlbertDevillard} we derive the asymptotic behavior of the WTD in the long time limit. For $R_A=1$ the decay is Gaussian with algebraic corrections well described by the Wigner surmise and for partial Andreev reflection ($R_A<1$) it is exponential with a rate given by the geometrical mean of the logarithm of $(1-R_A)$ over energy in $[\mu_S, \mu_S+eV]$, namely 
\begin{equation}
  \mathcal W(\tau)\simeq \exp\left[\overline{\ln(1-R_A)}\,\tau/\overline\tau\right]\, g(\tau/\overline\tau),
\end{equation}
where $g(y)$ is an oscillatory function that decays as $1/y^2$ and depends on $\overline{\ln^2(1-R_A)}$.

To conclude this section, we note that the role of electrons and holes may be interchanged by inverting the polarization ($eV\to -eV$) due to electron-hole symmetry.

\section{\label{sec:twospins}Effect of entanglement on waiting times}

\begin{figure}
  \includegraphics[width=0.9\linewidth]{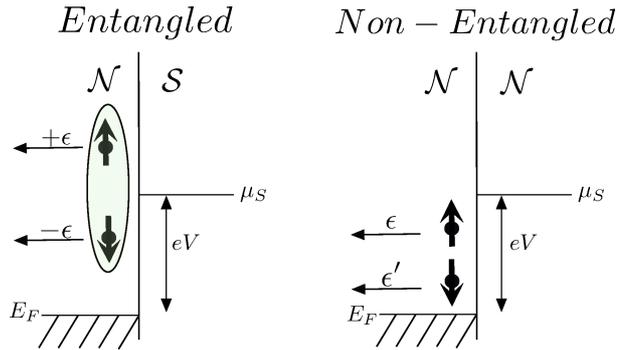}
  \caption{Schematic picture of the setup. Left: a Cooper pair is split at the interface of a NS junction and gives birth to two entangled electrons in the normal part. Right: two independent electrons as they would be emitted if the right part was also a normal metal.}
\label{fig:Fig3}
\end{figure}

We now move a step forward and discuss the effect of entanglement on waiting times. As mentioned before, Andreev reflection might be thought in terms of splitting of a Cooper pairs in the vicinity of the interface, leading to the injection of two entangled electrons with opposite spins and energy (with respect to $\mu_S$) in the normal part as shown in Fig. \ref{fig:Fig3} (left part). However, we have seen that the scattering state (Eq. (\ref{eq:psiout})) is a mixture of entangled and non-entangled components which makes the effect of entanglement hard to separate from other physical ingredients. Therefore, we focus here on the WTD of the fully entangled state of the form

\begin{equation}\label{eq:psiFE}
  \vert \psi_{{\rm FE}}\rangle \! =\!\!  \prod_{ E=0}^{eV} \left[\frac{c^{\dagger}_{k(E),\uparrow}c^{\dagger}_{-k(-E),\downarrow} -c^{\dagger}_{k(E),\downarrow}
c^{\dagger}_{-k(-E),\uparrow}}{\sqrt{2}}\right] \vert F \rangle\,,
\end{equation}
even if this is not the real quantum state emitted at the interface. Indeed, this work on NS junction rises a fundamental question of the effect of entanglement on waiting times that has never been considered to our knowledge and deserves to be investigated with proper care. At the end of this section we will give a few hints on how the additional terms included in the full state (Eq. (\ref{eq:psiout})) modify the picture but the common thread of this section will be the study of the fully entangled quantum state (Eq. (\ref{eq:psiFE})).  To point out the specific features of entanglement, we will compare this situation to the one of two independent electrons which would correspond to the emission of two electrons with opposite spins from a normal metal (see Fig. \ref{fig:Fig3} right) \cite{AlbertHaack14}. This time, the single electron detector is sensitive to both spins and energies between $\mu_S-eV$ and $\mu_S+eV$. Using the energy discretization mentioned before, the entangled state reads

\begin{equation}\label{eq:psiE}
  \vert \psi_{{\rm FE}}\rangle =  \prod_{n=1}^N \left[\frac{c^{\dagger}_{k_n,\uparrow}c^{\dagger}_{-k_n,\downarrow} -c^{\dagger}_{k_n,\downarrow}c^{\dagger}_{-k_n,\uparrow}}{\sqrt{2}}\right] \, \vert F \rangle\, ,
\end{equation}
whereas non-entangled electrons emitted from a normal lead in an energy window $eV$ (see Fig. \ref{fig:Fig3} right) would be described by $\vert \psi_{{\rm NE}}\rangle =  \prod_{n=1}^N c^{\dagger}_{-k_n,\uparrow} c^{\dagger}_{-k_n,\downarrow} \vert F \rangle$.

According to the detection process mentioned above, the ITP is formally given by \cite{Levitov96}

\begin{equation}\label{eq:idle2}
  \Pi(\tau) \, = \, \langle :e^{-Q_{\uparrow}}:\,:e^{-Q_{\downarrow}}: \rangle,
\end{equation}
where the quantum average is taken over the state $|\Psi_{{\rm FE}}\rangle$ or $|\Psi_{{\rm NE}}\rangle$ leading to $\Pi_{{\rm FE}}$ and $\Pi_{{\rm NE}}$ respectively. Without any further assumption it is clear that the final result cannot be expressed as a single determinant but rather as a sum of $2^{2N}$ terms except in the non-entangled case where the ITP factorizes to 

\begin{equation}
 \Pi_{\rm NE}(\tau)=\langle :e^{-Q_{\uparrow}}:\rangle\,\langle:e^{-Q_{\downarrow}}: \rangle=\det(1-Q_\uparrow) \det(1-Q_\downarrow) ,
\end{equation}
where the averages are taken over the two spin sectors separately \cite{AlbertHaack14}. Indeed, each term in the product of Eq. (\ref{eq:psiE}) can be split into two parts: ${\cal A}_n^{\dagger} = {c_{k_n,\uparrow}^{\dagger}c_{-k_n,\downarrow}^{\dagger} \over \sqrt{2}}$ and ${\cal B}_n^{\dagger} = - {c_{k_n,\downarrow}^{\dagger}c_{-k_n,\uparrow}^{\dagger} \over \sqrt{2}}$. $\vert \psi_{\rm FE}\rangle$ is then a sum of $2^N$ terms of the form $\prod_{i=1}^N {\cal C}_n^{\dagger} \vert F \rangle$, where ${\cal C}_n$ can be either ${\cal A}_n$ or ${\cal B}_n$. We shall denote a particular term by a string made of a succession of $A$'s and $B$'s defined as follows: $\vert A, B , ...,B \rangle \, \equiv \, {\cal A}_1^{\dagger}{\cal B}_2^{\dagger} ...{\cal B}_N^{\dagger} \vert F \rangle$. This is a Slater determinant, while $\vert \psi_{\rm FE} \rangle$ can not generally be cast as a simple Slater determinant. $\Pi_{\rm FE}$ will then be the sum of $2^{2N}$ terms of the form, typically, $T_{i,j} \equiv \langle \Psi_{S,i} \vert :e^{-Q_{\uparrow}}:\,:e^{-Q_{\downarrow}}: \vert \Psi_{S,j}\rangle$, with $\vert \Psi_{S,j} \rangle$ and  $\vert \Psi_{S,i} \rangle$ two generally different Slater determinants. As previously shown in Refs. \cite{Dasenbrook,Hassler08}, each $T_{i,j}$ is also a determinant. Details about the procedure to calculate the ITP are given in \ref{app:B}.

\subsection{Numerical results for the fully entangled case}

We now discuss our results obtained from a direct and exact enumeration of the $2^{2N}$ terms in the ITP. Owing to the exponentially growing number of terms, this approach is limited to relatively small values of $N$. Figure \ref{fig:Fig4}a) presents the WTD for increasing values of $N$ up to $N=12$. As can be seen, the curves reasonably converge to a limiting distribution that would be obtained for $N\to +\infty$. To insure this, we have computed several finite size corrections that we will discuss later on. Figure \ref{fig:Fig4}b) compares the WTDs of entangled and non-entangled electrons \cite{AlbertHaack14} which are qualitatively similar. However, as we will discuss in more detail in the next subsection, the maximum of the curve in the entangled case is more pronounced and closer to zero than in the independent case. The presence of such a peak in the WTD is the hallmark of pair rigidity due to entanglement. To be more quantitative we can evaluate the probability that the waiting time is smaller than the average waiting time $h/2eV$ (which is the same in both situations). We find that this probability is about ten percent larger in the entangled case, demonstrating that the entangled electrons are more correlated than the non-entangled ones. 

\subsection{Short time behavior for the fully entangled case}

The short time behavior is one of the most interesting characteristics of the WTD as it reflects the short time correlations encoded in the many body state and not the ones due to scattering. For free electrons it is universal since it is the expression of the Pauli's principle. Indeed, two spinless electrons cannot be emitted in the same state which enforces the WTD to start from zero (with a quadratic behavior). However, if the electrons have other degrees of freedom like spin or if the mesoscopic conductor supports several channels, the WTD may start from a non-zero value but correlations are still visible and universal \cite{AlbertHaack14}. However, these correlations only originate from electrons of the same channel and not between different channels. They eventually disappear in the limit of large number of independent channels \cite{AlbertHaack14}.

\begin{figure}
  \includegraphics[width=0.9\linewidth]{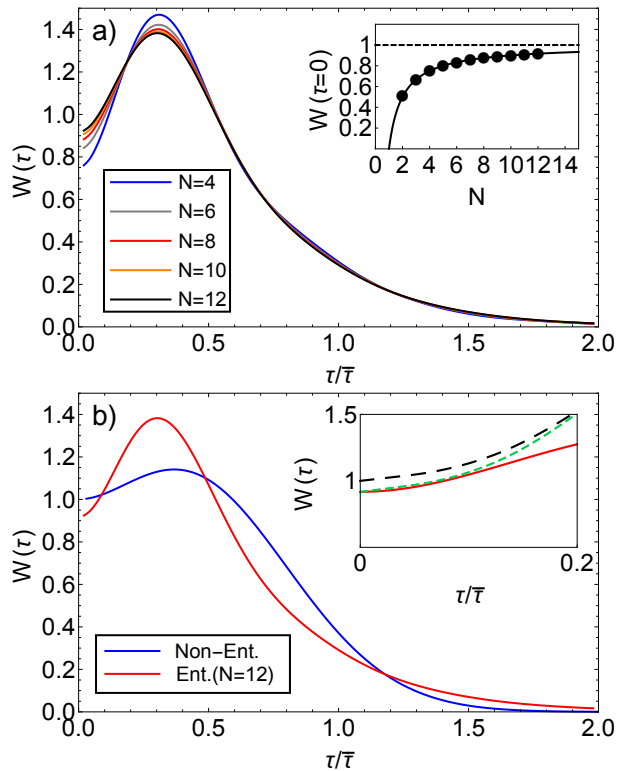}
  \caption{(Color online) WTD of fully entangled electrons. a) WTDs for increasing values of $N$. The inset shows how $\mathcal W(0)$ converges to the asymptotic value $1/\overline\tau$ as predicted by Eq. \ref{eq:stimewtd} (full black line). b) Comparison between the WTD of entangled electrons and non-entangled electrons (see text). Inset: short time behavior in the entangled case for $N=12$ (full red line and green dashed line (see Eq. \ref{eq:stimewtd})) and $N\to+\infty$ (black long dashed lines).}
\label{fig:Fig4}
\end{figure}

In order to get the short time expansion of the WTD up to second order in time, we need to expand the ITP (Eq. (\ref{eq:idle2})) to fourth order  in terms of moments of $Q_\downarrow$, $Q_\uparrow$ and their products. This is a straightforward but somehow cumbersome calculation that we do not reproduce in detail here. After some algebra we obtain the short time expansion of the WTD, including finite size corrections. The final expression for $\tau\ll\overline \tau$ reads, to second order in $\tau$

\begin{equation}\label{eq:stimewtd}
  \mathcal W(\tau)\simeq \frac{1}{\overline \tau}\left[\left(1-\frac{1}{N}\right)+C_N \frac{\pi^2}{3} \left(\frac{\tau}{\overline\tau}\right)^2\right]
\end{equation}
where $C_N=4(1+\frac{3}{2N})$ in the entangled case and one in the absence of NS junction. The insets of Fig.\ref{fig:Fig4}a) and Fig.\ref{fig:Fig4}b) show how this prediction is confirmed by our numerical evaluation of the WTD in the entangled case. In both situations, the WTD starts from the same initial value (in the $N\to \infty$ limit) with a quadratic behavior but the coefficient is four times larger in the entangled case. Again, this reflects the pair rigidity due to entanglement.

\subsection{Long time limit for the fully entangled case}

Differences between entangled and non-entangled electrons can also be inferred from the tail of the WTD. Following the approach developed in \cite{AlbertDevillard} we obtain the asymptotic behaviors of the WTD in the limit $\tau\gg \overline \tau$ (for details see \ref{app:B}). In both cases we find a Gaussian decay 

\begin{equation}
  \Pi(\tau)\sim\exp\left[-\frac{c}{2}\,\left(\frac{\tau}{\overline\tau}\right)^2\right]\,,
\end{equation}
with $c$ a constant which is equal to one in the entangled case and two in the non-entangled case. Therefore, the ITP or equivalently the WTD decays more slowly in the entangled case, meaning that the two electrons are not at all independent.

\subsection{Realistic situation in the tunneling limit}

As mentionned before, entangled pairs are only one component of the real out-going scattering state which appears to be much more complex. For abritrary value of the reflection coefficients $R_N$ and $R_A$, numerical evaluation of the WTD is pretty challenging since the number of terms grows as $4^{2N}$. In that case, direct numerical evaluations are restricted to very small values of $N$ (typically 4 or 5) which is not sufficient to mimic a stationary situation. It is then necessary to resort to more sophisticated approaches like Monte-Carlo sampling of the ITP for instance. However, there is a simple but rather interesting limit which is amenable to analytical calculations, namely the tunneling limit for $R_A\ll 1$ $(R_N \approx 1)$. In that case, the full many-body state simplifies to \cite{Samuelsson,Prada}

\begin{equation}
  \label{eq:psitunnel}
  |\psi_{\rm t}\rangle=\left[1+r_a r_n\sum_{E=0}^{eV}(c_{k,\uparrow}^\dagger c_{-k,\downarrow}^\dagger-c_{k,\downarrow}^\dagger c_{-k,\uparrow}^\dagger) \right]|F\rangle\,,
\end{equation}
namely, to first non-trivial order in $R_A$, it is made of scarce singlet pairs. Following the same approach as before we obtain the small time behavior
\begin{equation}
  \label{eq:psitunnelst}
  \mathcal W(\tau)\simeq 8\frac{R_A}{\overline \tau}\left[1-\frac{4}{3}\pi^2\left(\frac{\tau}{\overline \tau}\right)^2\right]
\end{equation}
and the long time asymptotics 
\begin{equation}
  \label{eq:psitunnellt}
  \mathcal W(\tau)\sim \exp[-R_A \tau/\overline \tau]\,.
\end{equation}

At short waiting times, the distribution starts from a constant value and the physics is dominated by correlations within a single Cooper pair. Indeed, simple considerations show that pairs are roughly separated in time by $\overline \tau/R_A$ whereas electrons from the same pair are rather in a span of time of the order of $\overline\tau$. The situation is then akin to a single pair which has been studied by Hassler et al. \cite{Hassler08} and confirms our predictions. At large time, the decay is exponential with a rate twice smaller than in the non-entangled case, which again is an hallmark of pair rigidity.

\section{\label{sec:conclusion}Conclusion}

We have presented a theory of waiting times in polarized NS junction at zero temperature. In this setup, Andreev reflection brings new characteristic features in the WTD such as entanglement. 

If a detector is sensitive to only one type of spin and to energies above the superconductor chemical potential, the situation is reminiscent of a single quantum channel connected to two normal leads via a QPC with energy-dependent transmission. The interface acts as an Andreev mirror and allows to measure the WTD of holes converted to electrons. However, if the detector measures electrons with both spins in the whole energy window above the Fermi sea of the normal conductor, entanglement between electrons leaves fingerprints in the WTD. Although still academical, we have focused on a fully entangled state which is only one component of the full many-body state flowing out of the NS interface. In that case, such signatures are visible for both small and large waiting times but the most important feature is the existence of a peak in the WTD centered before the average waiting time. When taking into account all components of the scattering state, we have shown that some charateristics are still visible in the WTD in the tunneling limit. 

In the near future, it would then be useful to extend this work beyond the ideal situation and the tunneling limit and evaluate the effect of entanglement for arbitrary values of reflection coefficients. In addition, a natural extensions would be to study correlations between waiting times of different spin species in the spirit of \cite{Hofer2015}. This would probably yield an even clearer signature of entanglement than the WTD itself.

Among other future investigations, it would be possible to study the effect of cross-Andreev reflection in a Superconducting-Normal-Superconducting junction or the physics brought by exotic states like Majorana modes created by Majorana guns \cite{Tarasinski,Gnezdilov}.

\section*{Acknowledgments}
The paper is dedicated to the memory of Markus B\"uttiker who was very enthusiastic about the quantum theory of waiting times. M. A., will always be grateful to him for his kindness, trust and freedom he gave to him when M. A. was a post-doc in his group. Markus was a mentor to him and a source of endless inspiration. We thank D. Dasenbrook, C. Flindt, G. Haack, P. P. Hofer and M. Moskalets for useful discussions and remarks. The research of D. C. was supported by the Foundation for Fundamental Research on Matter (FOM), the Netherlands Organization for scientific Research (NWO/OCW), and an ERC Synergy Grant.  


\appendix

\section{\label{app:A}Supplementay information on the Andreev mirror}
This appendix is dedicated to prove the formal correspondence between the WTD of a single channel normal conductor and the Andreev mirror defined in section \ref{sec:onespin}. This happens because the detection process is only sensitive to spin up electrons and energies above $\mu_S$, the superconducting chemical potential. To begin the proof we recall the definition of the ITP in terms of the scattering state

\begin{equation}
\Pi_{\uparrow}(\tau)=\langle \psi_{\rm out} \vert :e^{-Q_{\uparrow,k>k_S}}:\vert \psi_{\rm out} \rangle
\end{equation}
with 
\begin{align}
\vert \psi_{\rm out}\rangle\ &=\prod^{N}_{i=1}(r^*_N+r_A c^\dagger_{k_i,\uparrow}c^\dagger_{-k_i,\downarrow})(r_N-r^*_Ac^\dagger_{k_i,\downarrow}c^\dagger_{-k_i,\uparrow})\vert F\rangle\
\end{align}
where $N$ corresponds to the number of slices due to the energy discretization between $\mu_S$ and $\mu_S+eV$ (which has to be taken equal to infinity at the end of the calculation) and $k$ is measured with respect to $k_S$ the Fermi momentum at energy $\mu_S$. There are two important steps in the derivation. The first one consists in expanding the products and rearranging the terms in order make the relevant terms appear (the ones with spin up and positive energy electrons). If we restrict ourselves to these terms we build terms like $\vert{\cal C}\rangle=\prod^{N}_{i=1}(r^*_N+r_Ac^\dagger_{k_i,\uparrow}c^\dagger_{-k_i,\downarrow})\vert F \rangle$ . Then, we include the other terms of the product and show that they are irrelevant. Indeed, in the following example

\begin{align}
\langle {\cal C}\vert (r^{*}_N-r_A c_{-k_{i},\uparrow} c_{k_{i},\downarrow}) :e^{-Q_{\uparrow,k>k_S}}: (r_N-r^{*}_A c^\dagger_{k_{i},\downarrow}c^\dagger_{-k_{i},\uparrow}) \vert{\cal C} \rangle,
\end{align}
the two off-diagonal terms vanish since the operator $:e^{-Q_{\uparrow,k>0}}:$ conserves the number of particles. Concerning the two diagonal ones, they leave us with 

 \begin{equation}
 (\vert r_N\vert^2+\vert r_A\vert^2)\langle {\cal C}\vert :e^{-Q_{\uparrow,k>k_S}}:\vert{\cal C} \rangle= \langle {\cal C} \vert:e^{-Q_{\uparrow,k>k_S}}:\vert{\cal C}\rangle.
 \end{equation}
Then, we can repeat the same argumentation for all the other irrelevant terms from $i=1$ to $N$. Finally, we are left with

\begin{equation}
\langle F \vert \prod^N_{i=1}{\cal D}^\dagger_i :e^{-Q_{\uparrow,k>k_S}}: \prod^N_{j=1}{\cal D}_j\vert F \rangle
\end{equation}
with ${\cal D}_i= (r^{*}_N+r_Ac^\dagger_{k_i,\uparrow}c^\dagger_{-k_i,\downarrow})$. As a second step, we rearrange the Fock space by moving all the spin $\uparrow$ to the left. Spin $\downarrow$ are invariant under $:e^{-Q_{\uparrow,k>k_S}}:$, thus we only have to compute 

\begin{equation}
\langle F\vert \prod^N_{i=1} (r_N+r^{*}_A c_{k_i,\uparrow}) :e^{-Q_{\uparrow,k>k_S}}:\prod^N_{j=1}(r^{*}_N+r_Ac^\dagger_{k_j,\uparrow}) \vert F\rangle.
\end{equation}

 We then recognize the expression for the ITP of a spinless quantum channel with bias $eV$ and transmission amplitude $r_A$ \cite{Albert12}.

\section{\label{app:B}Fully entangled case}

\subsection{Numerical procedure in the fully entangled case}

In this Appendix we give some details on the construction on $\Pi(\tau)$. Starting from the N-body states reads
\begin{equation}\label{wave_function}
  \vert \psi \rangle\, = \prod^{N}_{n=1}\left[\frac{(c^{\dagger}_{k_n,\uparrow}c^{\dagger}_{-k_n,\downarrow} -c^{\dagger}_{k_n,\downarrow}c^{\dagger}_{-k_n,\uparrow})}{\sqrt{2}}\right] \, \vert F \rangle,
\end{equation}
with $\vert F \rangle$ is the Fermi sea defined in the main text and
\begin{equation}
k_n=\frac{2\pi n}{N}\frac{\vert eV\vert}{h v_F}.
\end{equation}
The probability to detect nothing in a range of time $\tau$ with a detector sensitive to all the energies and all spins can be written as
\begin{equation}\label{pitau}
\Pi(\tau)=\langle :e^{- Q_{\uparrow}}: :e^{- Q_{\downarrow}}: \rangle,
\end{equation}
where the average is taken over $\vert \psi \rangle$. From here, it is convenient to separate Eq. (\ref{wave_function}) in two parts $c^{\dagger}_{k_n,\uparrow}c^{\dagger}_{-k_n,\downarrow}$ and $c^{\dagger}_{k_n,\downarrow}c^{\dagger}_{-k_n,\uparrow}$ in order to get a sum of $2^N$ terms where each term is a Slater determinant. To simplify, it is useful to change the notation: now each configuration is associated to a specific ket (bra) and each element of this configuration can be mapped to an Ising classical spin. Namely, $c^{\dagger}_{k_n,\uparrow}c^{\dagger}_{-k_n,\downarrow}$ ($c^{\dagger}_{k_n,\downarrow}c^{\dagger}_{-k_n,\uparrow}$) is associated to $\sigma_i =+1 (-1)$. Eq. (\ref{pitau}) becomes \cite{Albert12,Hassler08}.
\begin{equation}
\Pi(\tau)=\frac{1}{2^{2N}}\sum_{\lbrace\sigma\rbrace_b,\lbrace\sigma\rbrace_k}\langle \lbrace\sigma\rbrace_b \vert M \vert\lbrace\sigma\rbrace_k\rangle\,,
\end{equation}
where the summation runs over the $2^{2N}$ configurations for the bra/ket (One example of possible configuration is $\langle \sigma_1 \sigma_2 \sigma_3\vert=\langle 1, -1,1\vert$ for a system consisting of three electron pairs). Each element of the matrix M can be written as \cite{Hassler08}
\begin{equation}
M_{n,m}=\delta_{n,m}-\frac{\textrm{sin}(\kappa_{n,m}\,X/N)}{\kappa_{n,m}}\textrm{exp}(i \kappa_{n,m}\,X/N)\,,
\end{equation}
with $\kappa_{n,m}=\sigma_n n-\sigma_m m$ and $X=\tau/\bar{\tau}$.

\subsection{Calculation of $\Pi(\tau)$ in the long time limit}

We now explain how to compute the long time asymptotics of the ITP of the fully entangled state. In that case, the matrix reduces to usual Toeplitz matrix which intervenes in the statistics of levels in the Gaussian Unitary Ensemble ensemble of random matrices \cite{Metha}. Namely, 
\begin{equation}
M = I - \exp(i \varphi_n) \sin(\varphi_n)/n\pi\,,
\end{equation} 
with $\varphi_n=n \pi X/N$ and $X=\tau/{\overline \tau}$ (except for an unimportant factor  $\pi$ in the angle $\varphi_n$). The asymptotic behavior of the WTD for large $\tau$ is equivalent to the statistics of having no level in a large energy range. Using this correspondence, the main terms read, for large  $\mathcal W(\tau) \simeq C (\tau/{\overline \tau})^{-1/4} \exp\lbrack-(\tau/{\overline \tau})^2/2\rbrack$,  with $C$ a rather complicated constant which can be find in reference \cite{Krasovsky}. The decay is Gaussian with algebraic corrections.

\end{document}